\documentclass[pre,aps,twocolumn,superscriptaddress,longbibliography]{revtex4-2}

\usepackage[english]{babel} 
\usepackage{graphicx}
\usepackage{amssymb,amsfonts,amsmath}
\usepackage{color}
\usepackage[hidelinks]{hyperref}

\usepackage[normalem]{ulem}

\usepackage{tikz}
\usetikzlibrary{shapes}
\usetikzlibrary{patterns}
\usetikzlibrary{angles, quotes}
\definecolor{bostonuniversityred}{rgb}{0.8, 0.0, 0.0}
\definecolor{chromeyellow}{rgb}{1.0, 0.65, 0.0}

\newcommand{\rv}{{\mathbf r}}
\renewcommand{\vec}{\mathbf}

\newcommand{\red}{\color{black}}

\usepackage{soul} 

\usepackage{amssymb}

\usepackage[final]{showlabels} 

\begin{document}


\title{
Routes to the density profile and structural inconsistency
}

\author{S.~M. Tschopp}
\affiliation{Department of Physics, University of Fribourg, CH-1700 Fribourg, Switzerland}
\author{H. Vahid}
\affiliation{Leibniz-Institute for Polymer Research, Institute Theory of Polymers, D-01069 Dresden, Germany}
\author{J.~M. Brader}
\affiliation{Department of Physics, University of Fribourg, CH-1700 Fribourg, Switzerland}
\date{\today}

\begin{abstract}
Classical density functional theory (DFT) is the primary method for investigations of inhomogeneous fluids in external fields. 
It requires the excess Helmholtz free energy functional as input to an Euler-Lagrange equation for the one-body density. 
A variant of this methodology, the force-DFT, uses instead the Yvon-Born-Green equation to generate density profiles. It is known that the latter are consistent with the virial route to the thermodynamics, while DFT is consistent with the compressibility route. In this work we will show an alternative DFT scheme using the Lovett-Mou-Buff-Wertheim (LMBW) equation to obtain density profiles, that are shown to be also consistent with the compressibility route.
However, force-DFT and LMBW DFT can both be implemented using a closure relation on the level of the two-body correlation functions.
This is proven to be an advantageous feature, opening the possibility of an optimisation scheme in which the structural inconsistency
between different routes to the density profile is minimized.
(`Structural inconsistency' is a generalization of the notion of thermodynamic inconsistency, familiar from bulk integral equation studies.)
Numerical results are given for the density profiles of two-dimensional systems of 
hard-core Yukawa particles with a repulsive or an attractive tail, in planar geometry.
\end{abstract}

\maketitle

\section{Introduction}

Density functional theory (DFT) provides an exact equilibrium framework for the study of classical many-body systems under the influence of external fields. 
The primary aim of DFT is to determine the one-body density profile and associated thermodynamic quantities by variational minimization of the grand potential functional. 
The non-trivial contribution to the grand potential is the excess (over ideal) Helmholtz free energy functional, 
$F^{\text{exc}}$, which encodes all information regarding the interparticle interactions and usually requires approximation. 

For simple model systems with purely hard-core repulsive 
interactions the best known approximations to $F^{\text{exc}}$ are 
given by fundamental measure theory (FMT). 
This approach has been shown to yield good results for many 
particle types of interest, such as spheres \cite{Rosenfeld89,Tarazona_Freezing,RothReview}, spherocylinders 
\cite{Wittmann_2015}, parallel cubes \cite{cuesta_cubes} and 
disks in two dimensions \cite{MartinDisks}. 
The situation is less satisfactory if the interaction potential 
is softly repulsive or has an attractive component (e.g.~the 
Lennard-Jones potential). 
The former is usually dealt with by mapping onto an effective hard particle system \cite{BarkerHendersonLiquid}, whereas the latter 
are typically treated using either the simplest mean-field approximation 
\cite{evans79,evans92} or other perturbative approaches 
\cite{MHNC_Rosenfeld,MHNC_Oettel,Tschopp1}.   

Given the challenges of directly approximating the free energy functional, it is not surprising that 
machine-learning (ML) methods are nowadays being exploited to 
either learn the free energy functional \cite{ML_Utrecht_2025_pair} 
or directly establish the connection between the one-body density and the direct correlation function \cite{ML_Bayreuth_2023,ML_Bayreuth_2024} -- the latter being the quantity required as input to the Euler-Lagrange (EL) equation for the density profile. Applications to specific systems, such as charged \cite{ML_Utrecht_2025_charged} or patchy \cite{ML_Tueb_2024_patchy,ML_Tueb_2025_neuralborgis} particles have proven quite successful. 
(For further details on machine learning in DFT we direct the reader to the recent review of Simon and Oettel \cite{ML_Tueb_2024_general}.)
While ML approaches can be effective for interpolating and extrapolating 
simulation data they retain a certain `black box' character which does not reveal to the user the fundamental mechanisms at work in 
the many-body system.
Consequently the desire to have a first-principles density functional scheme still remains.

It could be argued that the requirement of finding explicit analytic forms for $F^{\text{exc}}$ is, in fact, overly demanding and that this has limited the range of systems 
which can be addressed using standard DFT methodology. 
An alternative scheme, which relaxes this restriction, is provided by the integral equation theory of inhomogeneous liquids 
\cite{henderson92}. 
Recent work has revived and reinterpreted the inhomogeneous integral equation approach in a way which makes clear that this is indeed a DFT, but formulated on the level of two-body functionals \cite{tschopp8}.
This theory represents a natural generalization of previous work on bulk fluids.  
If one looks back into the development of liquid-state theory, from Kirkwood to the present day, then the majority of approaches were focused on bulk systems for which the pair correlation functions played a central role. Knowledge of these provides both detailed information about the internal microstructure of the fluid and, in the case of pair interactions, enables the calculation of all thermodynamic properties. 
There is no reason to suppose that the two-body correlations are less important for inhomogeneous systems and 
it seems that the only reason for their underuse is the computational complexity of their calculation.

Let us first consider the bulk pair correlations.
Thermodynamic quantities can be obtained from these via different routes, namely the compressibility and the virial.
If the pair correlation functions are known exactly, then both
routes are fully equivalent and yield identical results. However, since this is never the case for realistic 
model fluids, approximations are required, usually in the form 
of closures to the Ornstein-Zernike (OZ) integral equation \cite{hansen06,mcquarrie}. 
Predictions for thermodynamic 
quantities, such as the pressure, then become dependent on the chosen route; a phenomenon 
referred to as thermodynamic inconsistency 
\cite{caccamo,Janssen,BomontReview}.

Fortunately, thermodynamic inconsistency can be turned to 
our advantage by using it as a tool to generate improved 
approximation schemes. The development of thermodynamically 
consistent closures essentially began in 1973 with 
Waisman's generalized mean-spherical approximation, 
in which 
a free parameter in the closure relation was determined
by enforcing an accurate equation of state 
\cite{GMSA1}; an idea further developed by Stell and H\o ye in 
Reference \cite{GMSA2}. 
Subsequent focus was placed on the numerical determination of adjustable closure parameters to ensure consistency between 
the routes without imposing an equation of state as additional input to the theory. After the development of some classic 
consistent approximation schemes between 1979 and 1986
\cite{MHNC,verlet_closure,rogersyoung,Zerah,Ballone86}, the mid 
1990's saw a resurgence of interest \cite{scozaWilding,scozaOther,bigSCOZA,DuhHaymet,DuhHenderson,HRTreview,HRTandSCOZA,HRTrecent1,HRTrecent2,
Lee1,Lee2,SantosRational1,SantosRational2}. More recent attempts 
have largely been extensions/modifications of existing approximations to further improve accuracy, see e.g.~\cite{ModifiedVerletAttractive,Bomont,Carbajal-Tinoco}.

Turning now to inhomogeneous fluids we observe that there exist 
two independent expressions  
relating the inhomogeneous two-body correlation functions to 
the one-body density, namely
the Yvon-Born-Green (YBG) and
the Lovett-Mou-Buff-Wertheim (LMBW) sum-rules \cite{widom,JimHenderson92,AttardBook,Yvon_YBG,BornGreen_YBG,LMBW1,LMBW2,Lovett91,Lovett92}.
The YBG equation has been used by the present authors to study confined fluids in two-dimensions \cite{tschopp8}, but remains a relatively underexploited equation in studies of inhomogeneous fluids since its implementation presents certain technical challenges.
In contrast, the LMBW equation is easier to numerically implement and has thus received more attention. Some examples are the studies by Omelyan \textit{et al.} 
for the liquid-vapour interface \cite{kovalenko}, 
by Brader for pre-freezing 
phenomena \cite{Brader2008Struc-885} and by Nyg\aa rd 
\textit{et al.} for the prediction of local ordering in confined fluids \cite{nygard1,nygard2}.
Other works of interest are the study of hard-sphere solutes by Attard 
\cite{AttardSpherical,AttardSpherical2}
and the investigation of systems under planar confinement
by Kjellander and Sarman \cite{KjellanderSarman1,KjellanderSarman1}.

The YBG and LMBW expressions are formally equivalent, they however generally yield different one-body density profiles
if approximate two-body correlation functions 
are used as input. 
This sensitivity to the choice of sum-rule, which has recently been termed `structural inconsistency' \cite{tschopp8}, adds a layer of complexity and sublety 
to the study of inhomogeneous integral equations which 
has so far received very little attention.
Some clarity was gained in Reference \cite{tschopp3}, 
in which it was shown that 
the one-body density profiles obtained using the 
YBG equation are consistent with the virial route to 
the bulk thermodynamics. 
(This confirmed some earlier and long overlooked 
observations by Henderson \cite{henderson92}.) 
The situation may be less clear for the case of the 
LMBW equation in which the one-body density is generated by integrating over the two-body direct correlation function.

In this paper we develop a LMBW DFT scheme. We test the 
contact theorem, which relates the bulk pressure to an integral over the inhomogeneous density profile.
This reveals that the LMBW DFT generates compressibility consistent profiles. Standard DFT also yields compressibility consistent outcomes but, as already mentioned, always requires an expression for $F^{\text{exc}}$ as input. 
In contrast, both the YBG and the LMBW DFT schemes are implementable using a closure relation.
Since the YBG scheme yields virial results, we choose a closure relation involving a tuning parameter and vary it to enforce structural consistency between the two routes.
We show numerical results, in two-dimensional planar geometry, for hard-core Yukawa particles with a repulsive or an attractive tail.

\section{Methods}

\subsection{Bulk fluids}

\subsubsection{The bulk Ornstein-Zernike equation}\label{subsection bulk OZ}

In bulk the two-body correlation functions are solely dependent on the distance between two particle centers, $r_{12} \!\equiv\! |\vec{r}_1-\vec{r}_2|$. The total correlation function, $h$, and the two-body direct correlation, $c$, are linked via the bulk OZ equation \cite{hansen06,mcquarrie}
\begin{equation}\label{bulk OZ equation}
h(r_{12}) = c(r_{12}) + \rho_{\text{b}}\! \int \! d\vec{r}_3 \, h(r_{13})\, c(r_{32}),
\end{equation}
where $\rho_{\text{b}}$ is the bulk density. The radial distribution function, $g$, is related to $h$ according to $g\!\equiv\!h\!+\!1$.
Here and henceforth both volume integrals and densities have to be interpreted according to the spatial dimensionality 
under consideration.
A supplementary closure relation between $h$ and $c$ is required 
to obtain a closed system of equations. 

\subsubsection{Two routes to calculate the pressure}\label{subsection pressures}

If the exact two-body correlations are employed, then each 
of the different routes to calculate the thermodynamical quantities yields fully consistent results and the choice 
of one over the other is purely a matter of computational 
convenience. 
However, if the two-body correlations are only known approximately, as is almost always the case, then
the predictions from a given closure will vary from one choice of route to another.\\

\paragraph{The virial route:}

For a system in $D$ dimensions interacting 
via a pairwise additive potential, $\phi$, the virial pressure is given 
by
\begin{equation}\label{virial equation}
\frac{\beta P_{\text{v}}}{\rho_{\text{b}}} = 1 - \frac{\beta\rho_{\text{b}}}{2D} \int \! d\vec{r} \, r g(r) \frac{d\phi(r)}{dr} ,
\end{equation}
where $\beta\!\equiv\!(k_BT)^{-1}$. 
This result can be derived straightforwardly from the Clausius virial \cite{hansen06}.
For the two-dimensional systems to be addressed in the present 
work, equation \eqref{virial equation} yields
\begin{equation}\label{virial equation 2D}
\frac{\beta P_{\text{v}}}{\rho_{\text{b}}} \overset{2\text{D}}{=} 1 - \frac{\beta \rho_{\text{b}}}{2} \int_{0}^{\infty} \!\! dr \, r^2 g(r) \frac{d\phi(r)}{dr} .
\end{equation}
We note that if $\phi$ has a discontinuity (e.g.~for hard-core 
potentials), then one should be careful not to neglect the 
delta-function which occurs in its derivative. 
The virial expression \eqref{virial equation} has the appealing feature that the pressure can be evaluated directly by spatial integration of the radial distribution 
function and does not require an integration over thermodynamic parameters. \\

\paragraph{The compressibility route:}

The isothermal compressibility, $\chi_{T}$, is defined by the 
standard thermodynamic relation 
\begin{equation}\label{chiT definition}
\chi_{T} = \frac{1}{\rho_{\text{b}}}\left( \frac{\partial \rho_{\text{b}}}{\partial P} \right)_{T} \; .
\end{equation}
Alternatively, this can be expressed as an integral over
the direct correlation function according to \cite{hansen06}
\begin{equation}\label{compressibility equation}
\frac{\rho_{\text{b}} \chi_{T}}{\beta}
= \left( 1 - \rho_{\text{b}} \!\int \! d\vec{r} \,c(r) \right)^{-1} 
.
\end{equation}
Unlike the virial equation, this expression does not rely on the assumption of pairwise additivity of the interparticle potential. 
Introducing the static structure factor, $S$, 
provides the following equivalent, but more compact, expression 
\begin{equation}\label{compressibility equation fourier}
\frac{\rho_{\text{b}} \chi_{T}}{\beta}
= \left( 1 - \rho_{\text{b}} \, \tilde{c}(k\!=\!0) \right)^{-1}
\equiv S(k\!=\!0) ,
\end{equation}
where the tilde signifies a $D$-dimensional 
Fourier transform and $k$ is the magnitude of the 
wavevector.
Combining equations \eqref{chiT definition} and 
\eqref{compressibility equation fourier} and integrating 
over the bulk density generates
the familiar form for the compressibility pressure
\begin{equation}\label{compressibility pressure equation}
\frac{\beta P_{\text{c}}}{\rho_{\text{b}}} = 1 - \frac{1}{\rho_{\text{b}}} \int_{0}^{\rho_{\text{b}}} \!\! d\rho \, \rho \, \tilde{c}(k\!=\!0; \rho) ,
\end{equation}
where the dependence of the correlation functions on the density has been made explicit in the notation.

We observe that implementation of equation \eqref{compressibility pressure equation} is problematic 
for systems exhibiting a liquid-gas phase transition arising from 
a mutual attraction between the particles. At low temperatures 
the integration path to values of $\rho_{\text{b}}$ within the liquid 
phase will necessarily cross the spinodal region, where 
the direct correlation function appearing in equation 
\eqref{compressibility pressure equation} is not well-defined. The 
compressibility pressure at subcritical liquid states is 
thus not accessible, which then hinders the determination of a 
binodal.

\subsection{Inhomogeneous fluids}

\subsubsection{The inhomogeneous Ornstein-Zernike equation}

When an external potential, $V_{\text{ext}}$, is applied to the system then the symmetry of the bulk is broken.
This leads to a spatially varying one-body 
density, $\rho(\vec{r})$, and to two-body correlations which in general depend on two vector positions. 
The two-body direct correlation function and total correlation 
function are now related to each other by the inhomogeneous 
OZ equation \cite{evans92}
\begin{equation} \label{Inhomogeneous OZ equation}
h(\rv_1,\rv_2) = c(\rv_1,\rv_2) 
+ \!\int\!\! d\rv_3\, h(\rv_1,\rv_3)\,  \rho(\rv_3) \, c(\rv_3,\rv_2),
\end{equation}
which recovers equation \eqref{bulk OZ equation} in the bulk limit. 
In order to provide a closed framework for predictive calculation, 
equation \eqref{Inhomogeneous OZ equation} must be supplemented 
by: (i) a closure relation between $c$ and $h$ containing details of 
the interparticle interaction
and (ii) an exact sum-rule relating 
the inhomogeneous two-body correlations to the one-body density.  
Before considering approximate closures we will address the second 
point and discuss two distinct sum-rules (`routes') for 
the one-body density.

\subsubsection{Two routes to the one-body density}\label{subsection densities}

The equilibrium one-body density is 
defined by the following grand canonical average 
\cite{hansen06}
\begin{equation}\label{density_stat_def}
\rho(\textbf{r}_1) = 
\frac{1}{\Xi}\sum_{N=1}^{\infty} 
\frac{e^{\beta\mu N}}{(N-1)!}
\int\! d\textbf{r}_2\cdots\!\!\int\! d\vec{r}_N 
e^{-\beta U_N},
\end{equation}
where $\Xi$ is the grand canonical partition function, 
$\mu$ is the chemical potential and 
$U_N$ is the total potential energy. 
For systems interacting via pairwise additive potentials the 
latter is given by
\begin{equation}\label{tot potential}
U_N(\{\vec{r}^N\}) = 
\sum_{i=1}^{N}V_{\text{ext}}(\vec{r}_i) 
+
\frac{1}{2}\sum_{\substack{i,j=1\\i\ne j}}^{N}
\phi(r_{ij}).
\end{equation}
The inhomogeneous two-body density is defined by
\begin{equation}\label{twobodydensity_stat_def}
\!\!\!\rho^{(2)}(\vec{r}_1,\vec{r}_2) = 
\frac{1}{\Xi}\sum_{N=2}^{\infty} 
\frac{e^{\beta\mu N}}{(N-2)!}
\int\!\! d\vec{r}_3 \! \cdots\!\!\int\!\! d\vec{r}_N 
e^{-\beta U_N}\!.
\end{equation}
Both the one- and two-body densites are explicitly functionals 
of the external potential.
Taking the functional derivative of equation \eqref{density_stat_def} generates the first Yvon equation \cite{Yvon,evans79}
\begin{equation}\label{first_yvon}
\!\!\frac{\delta \rho(\vec{r}_1)}{\delta \beta V_{\text{ext}}(\vec{r}_2)}
=
-\rho(\vec{r}_1)\rho(\vec{r}_2)
h(\vec{r}_1,\vec{r}_2)
-\rho(\vec{r}_2)\delta(\vec{r}_1-\vec{r}_2),
\end{equation}
where the total correlation function is related to the two-body 
density according to
\begin{equation}\label{link between h and rho2}
h(\vec{r}_1,\vec{r}_2) = 
\frac{\rho^{(2)}(\vec{r}_1,\vec{r}_2)}
{\rho(\vec{r}_1)\rho(\vec{r}_2)} - 1.
\end{equation}
Since the one-body density and the external potential must satisfy the following
functional chain-rule identity
\begin{equation}\label{chain rule}
\int d\vec{r}_3 \frac{\delta \rho(\vec{r}_1)}
{\delta V_{\text{ext}}(\vec{r}_3)}
\frac{\delta V_{\text{ext}}(\vec{r}_3)}
{\delta \rho(\vec{r}_2)}
=
\delta(\vec{r}_1-\vec{r}_2)
\end{equation}
and, since we have already stated equations \eqref{Inhomogeneous OZ equation} and \eqref{first_yvon}, it follows that the inverse of the latter must be given by
\begin{equation}\label{second_yvon}
\frac{\delta \beta V_{\text{ext}}(\vec{r}_1)}
{\delta \rho(\vec{r}_2)}
= c(\vec{r}_1,\vec{r}_2) 
- \frac{1}{\rho(\vec{r}_1)}\delta(\vec{r}_1-\vec{r}_2), 
\end{equation}
which is known as the second Yvon equation \cite{Yvon,evans79}.
Note that equation \eqref{second_yvon} essentially defines the inhomogeneous two-body direct correlation function \cite{evans79,evans92,widom}.\\

\paragraph{The Yvon-Born-Green equation:}\label{subsubsection about YBG}

Taking the gradient of the one-body density 
\eqref{density_stat_def} generates
\begin{align}\label{gradient_rho}
&\,\nabla_{\vec{r}_1}  \rho(\vec{r}_1) = 
-\rho(\vec{r}_1) \nabla_{\vec{r}_1} \beta V_{\text{ext}}(\vec{r}_1) 
\\ 
&-\frac{\beta}{\Xi}\sum_{N=2}^{\infty} 
\frac{e^{\beta\mu N}}{(N-2)!}
\!\int\! d \vec{r}_2 \,  
\nabla_{\vec{r}_1} \phi(r_{12})
\int\! d\vec{r}_3\cdots\!\!\int\! d\vec{r}_N 
e^{-\beta U_N}
\!. 
\notag
\end{align}
Substituting the definition of the two-body density \eqref{twobodydensity_stat_def} into expression \eqref{gradient_rho} then yields the first YBG equation
\begin{align}\label{YBG equation}
\nabla_{\vec{r}_1}  \rho(\vec{r}_1) &=  -\rho(\vec{r}_1) \nabla_{\vec{r}_1} \beta V_{\text{ext}}(\vec{r}_1) \\
&- \!\int\! d \vec{r}_2 \,  \rho^{(2)}(\vec{r}_1,\vec{r}_2) \nabla_{\vec{r}_1} \beta \phi(r_{12}). \notag
\end{align}
This sum-rule was derived long ago 
by Yvon \cite{Yvon_YBG} and Born and Green \cite{BornGreen_YBG} and is a fundamental relation in 
statistical physics. 
Although we show here the standard derivation of equation 
\eqref{YBG equation} we note that an instructive alternative derivation 
can be found in Reference \cite{Lovett92}.
The YBG equation expresses the equilibrium force-balance between Brownian, external and internal forces, where the latter are intuitively expressed as a statistical average of the pair interaction force acting at the point $\vec{r}_1$. 
The closed framework generated by the YBG equation \eqref{YBG equation}, the inhomogeneous OZ equation \eqref{Inhomogeneous OZ equation} and an additionnal closure relation forms the force-DFT approach \cite{tschopp3,tschopp8}, which provides an alternative to 
the standard (variational) implementation of classical DFT 
\cite{evans92,evansKazimierz_Notes_2009}.\\

\paragraph{The Lovett-Mou-Buff-Wertheim equation:}\label{subsubsection LMBW}

Taylor expanding the external potential about 
the point $\vec{r}_1$ generates the following expression
\begin{align}\label{LMBW_first}
&V_{\text{ext}}(\vec{r}_1+\vec{r}') 
- V_{\text{ext}}(\vec{r}_1)
\\&\qquad=
\int\! d\vec{r}_2 
\frac{\delta V_{\text{ext}}(\vec{r}_1)}{\delta\rho(\vec{r}_2\!+\!\vec{r}')}
\big(
\rho(\vec{r}_2 + \vec{r}') - \rho(\vec{r}_2)
\big).\notag
\end{align}
If the points $\vec{r}_1$ and $\vec{r}'$ lie close together, such that 
$d\vec{r}'\!=\!\vec{r}_1-\vec{r}'$ is small in magnitude, 
then we can rewrite equation \eqref{LMBW_first}
using gradients, yielding
\begin{equation}\label{gradient_relation_inverse}
\nabla_{\vec{r}_1} V_{\text{ext}}(\vec{r}_1)
= \int d\vec{r}_2
\frac{\delta V_{\text{ext}}(\vec{r}_1)}{\delta \rho(\vec{r}_2)}
\nabla_{\vec{r}_2}\rho(\vec{r}_2).
\end{equation}
Substitution of the second Yvon equation \eqref{second_yvon} into expression
\eqref{gradient_relation_inverse} then leads to the 
 LMBW equation
\cite{LMBW1,LMBW2}
\begin{align}\label{LMBW}
\nabla_{\vec{r}_1}  \rho(\vec{r}_1) &=  -\rho(\vec{r}_1) \nabla_{\vec{r}_1} \beta V_{\text{ext}}(\vec{r}_1) \\
&+ 
\rho(\vec{r}_1)\!\int\! d \vec{r}_2 \,  c(\vec{r}_1,\vec{r}_2) \nabla_{\vec{r}_2} 
\rho(\vec{r}_2). \notag
\end{align}
For further insights we direct the reader to the papers 
of Lovett and Buff \cite{Lovett91} and Baus and Lovett 
\cite{Lovett92} as well as the excellent book by Widom and 
Rowlinson \cite{widom}.

\subsubsection{Closure relation approximations}

Using the indirect correlation function, $\gamma\!\equiv\!h-c$, the Verlet (V) and the Modified Verlet (MV) closure relations \cite{Janssen,verlet_closure} impose that $c(\vec{r}_1, \vec{r}_2)$ is given by
\begin{align}
&\,\overset{\text{V}}{=} e^{\!\!-\beta \phi(r_{12}) + \gamma(\vec{r}_1, \vec{r}_2)- \frac{\frac{1}{2} \gamma^2(\vec{r}_1, \vec{r}_2)}{1+\frac{4}{5} \gamma(\vec{r}_1, \vec{r}_2)}} - \gamma(\vec{r}_1, \vec{r}_2) \!-\! 1 , \label{V closure}\\
&\overset{\text{MV}}{=}e^{\!\!-\beta \phi(r_{12}) + \gamma(\vec{r}_1, \vec{r}_2)- \frac{\frac{1}{2} \gamma^2(\vec{r}_1, \vec{r}_2)}{1+\alpha_{\text{V}} \gamma(\vec{r}_1, \vec{r}_2)}} - \gamma(\vec{r}_1, \vec{r}_2) - 1,\label{MV closure}
\end{align}
respectively. The tuning parameter $\alpha_{\text{V}}$ in \eqref{MV closure} is zero for $\gamma\!<\!0$.
The Verlet closure is known to perform very well not only in bulk systems \cite{Janssen}, but also in the inhomogeneous case.
Indeed, in the recent paper \cite{tschopp8}, the Verlet closure was tested against other closure relations, namely the Hypernetted chain, the Percus-Yevick and the Martinov-Sarkisov ones, for two-dimensional inhomogeneous fluids and proved very successful.
In the present work we will thus use the Verlet closure in both original and modified versions.

\subsubsection{Wall contact theorem}\label{subsection contact thm}

Let us consider a two-dimensional planar wall represented by the external potential $V_{\text{ext}}(z)$, with the $z$-axes taken perpendicular to the wall.
Since the one-body density always exhibits the same symmetry as the external potential \cite{evans79}, it reduces to $\rho(\vec{r})\!=\!\rho(z)$.
For a generic wall potential, the following expression
\begin{equation}\label{general wall theorem}
\beta P = 
- 
\int_{-\infty}^{+\infty}dz\, \rho(z)
\left(\frac{d\beta V_{\text{ext}}(z)}{dz}\right)
\end{equation}
holds.
This is the general form of the so-called wall contact theorem
that relates the bulk pressure, $P$, to the forces acting on the wall, making a link between the bulk equations of state discussed in subsection \ref{subsection pressures} and the sum-rules for the inhomogeneous one-body density of subsection \ref{subsection densities}.

For an external potential consisting of a hard wall plus an either repulsive or attractive tail, 
$V_{\text{ext}}^{\text{tail}}(z)$, the derivative appearing in equation \eqref{general wall theorem} 
picks up a delta-function contribution. In this case, the contact theorem is expressed as
\begin{equation}\label{specific wall theorem}
\beta P = \rho(z_\text{w}) 
- 
\int_{z_{\text{w}}}^{+\infty}dz\, \rho(z)
\left(\frac{d\beta V_{\text{ext}}^{\text{tail}}(z)}{dz}\right),
\end{equation}
where $z_{\text{w}}$ is the $z$-coordinate of a particle in contact with the (hard) wall 
\cite{JimHenderson92}.

\subsection{Numerical implementation}

\subsubsection{System of interest: specification of the interparticle and external potentials}

\paragraph{Choice of the interparticle pair potential:}

In this work we will consider a two-dimensional system of 
hard-core Yukawa (HCY) particles, given by the following pair potential
\begin{align}\label{HCY potential}
\phi(r_{12}) = 
\begin{cases} 
      \hspace*{0.9cm}\infty, & r_{12} < d, \vspace*{0.2cm}\\
      \kappa\,\frac{e^{-\alpha(r_{12}-d)}}{r_{12}}, & r_{12}\ge d,
\end{cases}
\end{align}
where we set the particle (hard-core) diameter $d\!=\!1$.
We fix the inverse decay length $\alpha\!=\!2$ and we will vary theamplitude, $\kappa$, to cover physically different particle systems.
Imposing a positive value of $\kappa$ yields a fully repulsive pair potential, while changing its sign to a negative $\kappa$-value yields an attractive tail and thus a pair potential that is both repulsive and attractive. The latter can undergo phase separation which adds an extra layer of complexity to the problem. On the other hand increasing or decreasing the (always positive) value of $\alpha$ would lead to a shrinking or a broadening of the tail-range. In the limit of $\kappa\!\rightarrow\!0$, or equivalently of $\alpha\!\rightarrow\!\infty$, we recover a purely hard-core pair potential, namely hard disks.\\

\paragraph{Choice of external potentials:}\label{choice of ext pot}

\begin{figure}[t!]
\includegraphics[width=0.8\linewidth]{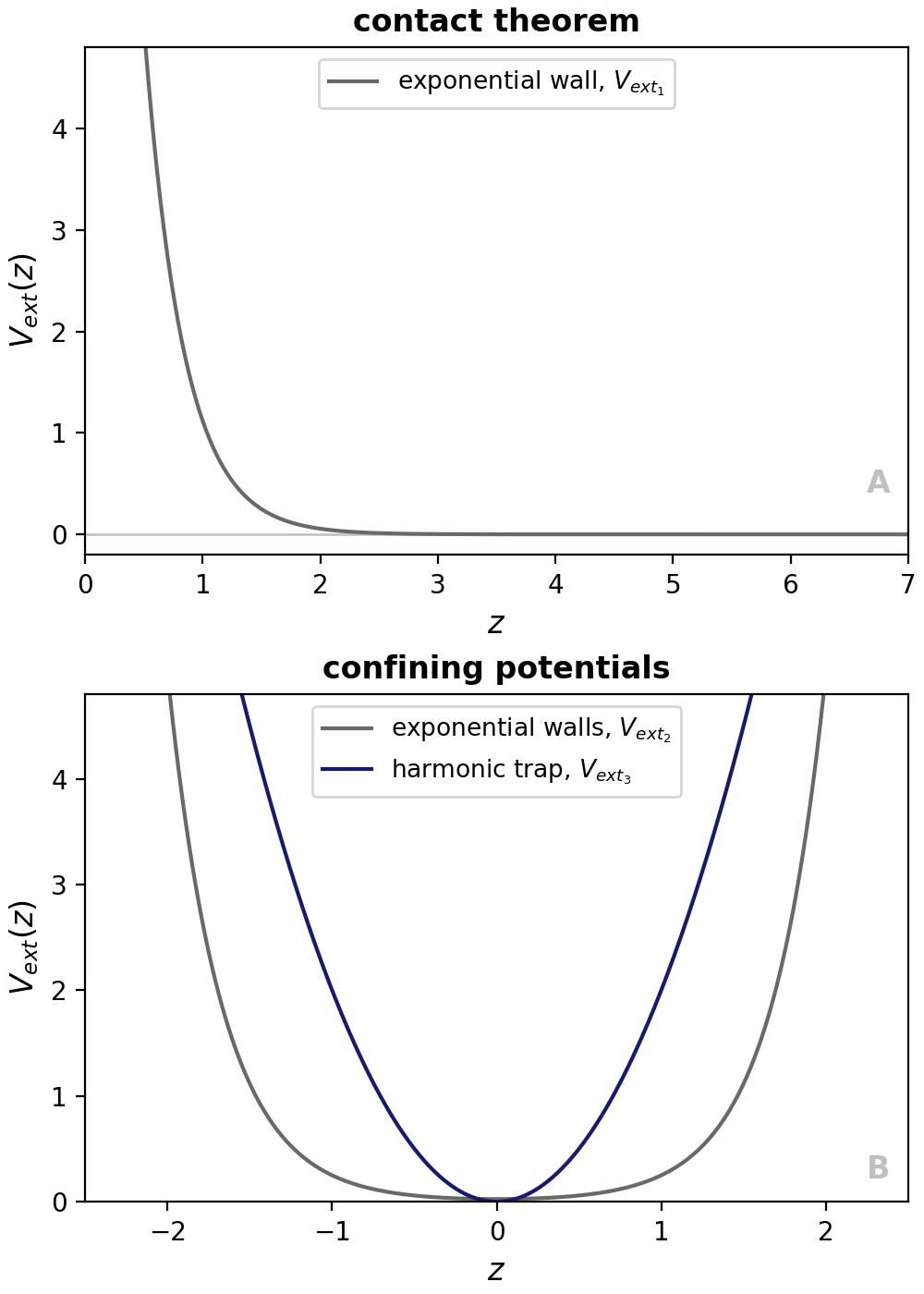}
\caption{ \textbf{Selection of external potentials.}
The softened repulsive hard-wall potential \eqref{external potential contact thm} is used to test the contact theorem. Its exponential tail is shown in panel A.
The soft parts of both confining external potentials \eqref{external potential exp walls} and \eqref{external potential harmonic trap} are shown in panel B.
} 
\label{fig choice Vext}
\end{figure}

In the inhomogeneous calculations for the contact theorem, we will use a softened repulsive hard-wall potential,
\begin{align}\label{external potential contact thm}
\! V_{\text{ext}_{1}}(z) = 
\begin{cases}
      \tilde{\kappa}\, e^{-\tilde{\alpha} (z - \frac{d}{2})}, & \!\! z > \frac{d}{2}, \vspace*{0.2cm}\\
      \hspace*{0.9cm}\infty, & \!\text{otherwise},
\end{cases}
\end{align}
i.e. a hard wall located at position $z\!=\!0$ plus an additional exponential soft-repulsive tail, with fixed amplitude $\tilde{\kappa}\!=\!5$ and inverse decay length $\tilde{\alpha}\!=\!3$, as illustrated in panel A of Figure~\ref{fig choice Vext}.

We will then proceed to consider particles confined within slits. The external potential is first chosen as
\begin{align}\label{external potential exp walls}
\!\! V_{\text{ext}_{2}}(z) = 
\begin{cases}
      \tilde{\kappa} \left( e^{\tilde{\alpha} (z - z_{w})} + e^{-\tilde{\alpha} (z + z_{w})} \right), & \!\! |z| < z_{w}, \vspace*{0.2cm}\\
      \hspace*{0.9cm}\infty, & \!\text{otherwise},
\end{cases}
\end{align}
where the hard walls are located at positions $z\!=\!\pm z_{w}$, with $z_{w} \equiv \frac{L}{2}\!-\!\frac{d}{2}$ and slit width {\red $L\!=\!5d$}. We set again the amplitude $\tilde{\kappa}\!=\!5$ and the inverse decay length $\tilde{\alpha}\!=\!3$, as parameters for the soft-repulsive tail.
Finally, we focus on a truncated harmonic trap given by
\begin{align}\label{external potential harmonic trap}
\! V_{\text{ext}_{3}}(z) = 
\begin{cases}
      \mathcal{A} \left( z-z_0 \right)^2, & \!\! |z| < z_{w}, \vspace*{0.2cm}\\
      \hspace*{0.9cm}\infty, & \!\text{otherwise},
\end{cases}
\end{align}
where $\mathcal{A}\!=\!2$ is the amplitude and $z_0\!=\!0$ the center of the trap.
Both confining potentials are shown in panel B of Figure~\ref{fig choice Vext}.

Given such external potentials, we already know from subsection \ref{subsection contact thm} that the one-body density reduces to $\rho(z)$.
Focusing henceforth on two-dimensional systems, we will use the cartesian coordinates $x$ and $z$.
Since the one-body density does not vary in the $x$-direction and since we can freely choose to fix the coordinate axes such that $x_1\!=\!0$, the two-body correlations are fully defined by $z_1$, $z_2$ and $x_2$.

\subsubsection{Obtaining the inhomogeneous density profiles}

As introduced in subsubsection \ref{subsubsection about YBG}, solution of the closed framework consisting of the YBG equation \eqref{YBG equation}, the inhomogeneous OZ equation \eqref{Inhomogeneous OZ equation} and an additionnal closure relation yields the force-DFT formalism.
Its implementation in two-dimensional planar geometry is given in detail in Reference \cite{tschopp8} and summarised below.
A natural generalization of this scheme is to perform a force-DFT-like resolution of the same closed framework, but using the LMBW equation \eqref{LMBW} in place of the YBG equation. (For clarity we will henceforth refer to force-DFT as YBG DFT.)\\

\paragraph{YBG and LMBW DFT in two-dimensional planar geometry:}

For two-dimensional planar geometry the YBG \eqref{YBG equation} and LMBW \eqref{LMBW} equations reduce to
\begin{equation}\label{compact equations planar}
\frac{d\rho(z_1)}{d z_1}  = - \rho(z_1) \frac{d\beta V_{\text{ext}}(z_1 )}{d z_1} + A(z_1),
\end{equation}
where $A$ is given by
\begin{align*}
& \!\! \overset{\text{YBG}}{=} - \int_{-\infty}^{+\infty} \!\! dx_2 \int_{-\infty}^{+\infty} \!\! dz_2 \, \rho^{(2)}(z_1,x_2,z_2 ) \frac{d\phi(r_{12})}{d z_1}, \\
& \!\!\!\!\! \overset{\text{LMBW}}{=} \rho(z_1) \int_{-\infty}^{+\infty} \!\! dx_2 \int_{-\infty}^{+\infty} \!\! dz_2 \, c(z_1,x_2,z_2 ) \frac{d\rho(z_2)}{d z_2},
\end{align*}
respectively, with the distance between the particles $r_{12}\!=\!\sqrt{x_2^2 + (z_1-z_2)^2}$
{\red and we recall that the particle diameter is set to unity.}
This yields the following equation for the one-body density
\begin{align}\label{EL-like equation}
\rho(z) &= \rho_{0} \, e^{-\beta V_{\text{ext}}(z)} \rho_{\text{exp}}(z),
\end{align}
where
\begin{align}
\rho_{\text{exp}}(z) \equiv e^{\int_0^z dz_1 \frac{1}{\rho(z_1)} A(z_1) }
\end{align}
and the integration constant is given by
\begin{equation}\label{force-DFT normalization}
\rho_{0} \!\equiv\! \frac{\langle N \rangle}{\int_{-\infty}^{+\infty} \! dz \, e^{-\beta V_{\text{ext}}(z)} \rho_{\text{exp}}(z)}, 
\end{equation}
for a given average number of particles per unit length, $\langle N \rangle\!=\!\int_{-\infty}^{+\infty} dz \,\rho(z)$.\\

\paragraph{Accounting for discontinuities:}

When implementing the YBG and LMBW versions of equation \eqref{EL-like equation} care must be taken for discontinuous pair potentials and density profiles, respectively, since their derivatives are present in $A$.

If one wishes to treat a system of hard-core particles, $A(z_1)$ for the YBG calculation is given by
\begin{align*}
& - \int_{-\infty}^{+\infty} \!\! dx_2\int_{-\infty}^{+\infty} \!\! dz_2 \, \rho^{(2)}(z_1,x_2,z_2 ) \frac{d\phi(r_{12})}{d z_1} \\
&= {\red - \int_{z_1-1}^{z_1+1} \!\! dz_2 \, \sqrt{1-(z_1-z_2)^2} \left(\frac{d}{dz_2}\rho^{(2)}(z_1,x_2^\star,z_2 )\right) } ,\\
\end{align*}
where {\red $x_2^\star\!\equiv\!\sqrt{1-(z_1-z_2)^2}$}.

On the other hand, if one uses hard walls as an external potential, the one-body density will be discontinuous at those wall-positions and thus $d \rho(z_2)/dz_2$ will yield a delta-function in the calculus of $A$ for the LMBW scheme. For a slit, the integral over $z_2$ would then become
\begin{align*}
\int_{-\infty}^{+\infty} &\!\!\! dz_2  \, c(z_1,x_2,z_2 ) \frac{d \rho(z_2)}{d z_2}\\
&= \int_{z_{\text{wall}}^{\text{left}}}^{z_{\text{wall}}^{\text{right}}} \!\!\! dz_2 \, c(z_1,x_2,z_2 ) \frac{d \rho(z_2)}{d z_2}
\\&\quad+ {\big(c(z_1,x_2,z_2 ) \rho(z_2) \big)}\big\rvert_{z_2=z_{\text{wall}}^{\text{left}}}
\\&\quad- {\big(c(z_1,x_2,z_2 ) \rho(z_2) \big)}\big\rvert_{z_2=z_{\text{wall}}^{\text{right}}} \; . \\
\end{align*}

For the detailed calculations of those two special cases, we refer the reader to Appendix~\ref{appendix accounting for discontinuities}.\\

\paragraph{The inhomogeneous OZ equation in two-dimensional planar geometry:}

When implementing the inhomogeneous OZ equation \eqref{Inhomogeneous OZ equation} in two-dimensional planar geometry, we use a one-dimensional Fourier transform in the $x$-direction, since the density only varies with respect to the coordinate $z$. We thus obtain
\begin{align} \label{FT OZ equation planar eq1}
\tilde{h}(z_1,k,z_2) &= \tilde{c}(z_1,k,z_2) \\
&\quad+ \!\int_{-\infty}^{+\infty}\!\! dz_3\, \tilde{h}(z_1,k,z_3)\,  \rho(z_3) \, \tilde{c}(z_3,k,z_2), \notag
\end{align}
where $\tilde{h}$ and $\tilde{c}$ are the Fourier transforms of $h$ and $c$ with respect to the $x_2$-coordinate.
Furthermore, we employ the (always spatially continuous) indirect correlation function, $\gamma$, to rewrite the equation \eqref{FT OZ equation planar eq1} as
\begin{align} \label{FT OZ equation planar eq2}
\tilde{\gamma}(z_1,k,z_2) &= \!\int_{-\infty}^{+\infty}\!\! dz_3\, \tilde{c}(z_1,k,z_3)\,  \rho(z_3) \, \tilde{c}(z_3,k,z_2) \\
&\quad+\!\int_{-\infty}^{+\infty}\!\! dz_3\, \tilde{\gamma}(z_1,k,z_3)\,  \rho(z_3) \, \tilde{c}(z_3,k,z_2). \notag
\end{align}
{\red In the present work numerical evaluation of the Fourier transform is performed using the method of Lado \cite{Lado}. The spacing of the gridpoints in both the $x$- and $z$-directions are (approximately) set to $dx\!\approx\!dz\!=\!0.05$, such that the two-body correlation matices can still be supported by the working memory of our workstation. The spatial cutoff value in the $x$-direction has been adapted in each presented case to ensure the best accuracy and stability, since long-range transverse correlations appear at high packing. Typical cutoff values are about $20$ particle diameters. We note that the computational running time increases rapidly with the number of gridpoints.}

\subsubsection{Generating simulation data for comparison}\label{sim data main}

In order to compare our numerical outputs with simulation data, we perform, in addition, overdamped
Brownian-dynamics (BD) simulations of HCY particles using \textsc{LAMMPS} package~\cite{LAMMPS2022}. The system is two-dimensional in the $xz$-plane with periodic boundary conditions along the $x$-axis. Confinement is imposed by an external potential that diverges outside a finite slab located inside the simulation box (slit geometry). Consequently, the height of the box, $L_z$, exceeds the width of the accessible slit and the particles cannot reach the regions where the external potential is effectively infinite. We use a rectangular box of size $L_x \times L_z = 1000d \times 12d$. For each choice of external potential (see subsubsection~\ref{choice of ext pot}), particles of diameter $d\!=\!1$ are initialized at random, without overlap, within the accessible slit region.
Pairs of particles interact via a steep, purely repulsive Mie potential with exponents $100/99$, as well as the Yukawa contribution (see equation~\eqref{HCY potential}). The particle positions $\mathbf r_i\!=\!(x_i,z_i)$ evolve according to $\gamma\,\dot{\mathbf r}_i(t)=\boldsymbol{\xi}_i(t)-\nabla_i U_N,$
where $U_N$ is given by equation \eqref{tot potential}, $\gamma$ is the translational friction coefficient set to unity, and $\boldsymbol{\xi}_i(t)$ is a Gaussian random force. Each simulation is conducted for $5\times 10^{8}$ steps with a time step $10^{-5}$. The initial $1\times 10^{8}$ steps are discarded as equilibration. Data are recorded every $2\times 10^{3}$ steps, and all results are reported from the post-equilibration part of the simulations.
For more specific details on each of the simulation curves presented in the results section, we refer the reader to Appendix~\ref{appendix sim data details}.

\section{Results}

\subsection{Contact theorem}

\begin{figure}[t!]
\includegraphics[width=1\linewidth]{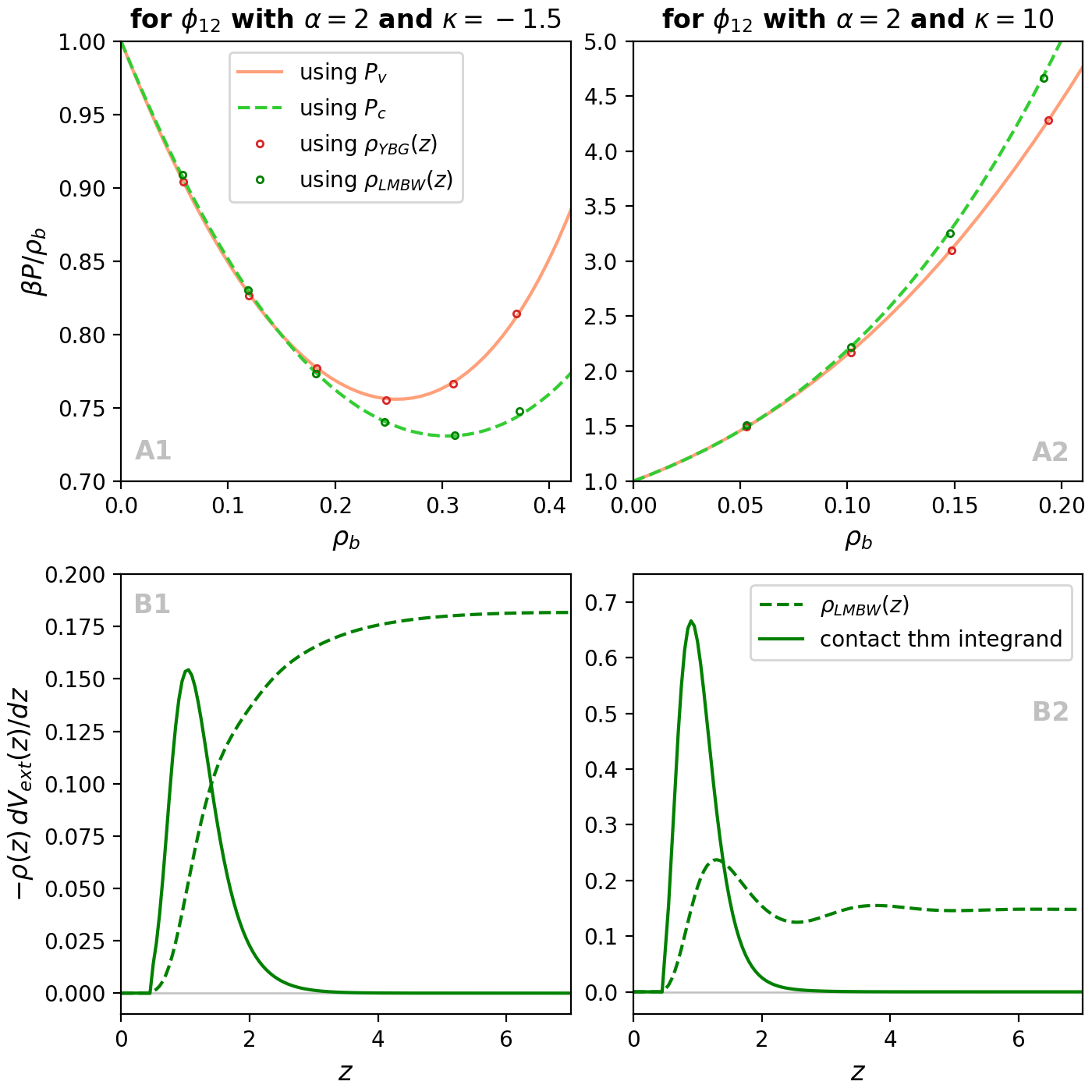}
\caption{ \textbf{Testing the contact theorem.}
We consider two types of HCY particles \eqref{HCY potential}, both with $\alpha\!=\!2$. Column 1, shows results for $\kappa\!=\!-1.5\!<\!0$, thus an attractive tail. Column 2, shows results for particles with a repulsive tail, with $\kappa\!=\!10\!>\!0$.
In panels A, we show reduced pressure curves calculated using both the virial \eqref{virial equation 2D} 
and compressibility \eqref{compressibility pressure equation} equations of state, given by the solid orange and dashed limegreen curves, 
respectively. 
The red circles show the results obtained via equation \eqref{specific wall theorem}, when the input density profile is generated by the 
YBG equation \eqref{YBG equation}.
The green circles show the same, when the input density profile is generated by the LMBW equation \eqref{LMBW}.
In panels B, we show the contact theorem integrand as solid green lines and the density profile generated by the LMBW equation as dashed green lines, to illustrate how we obtain the circles shown in panels A.
} 
\label{fig contact thm}
\end{figure}

As a first numerical investigation we employ the HCY potential \eqref{HCY potential} with the wall potential 
\eqref{external potential contact thm} to calculate density profiles using both the YBG equation \eqref{YBG equation} and 
the LMBW equation \eqref{LMBW}. In each case we use the wall contact theorem \eqref{specific wall theorem} to calculate the 
bulk pressure from our numerically obtained density profiles and compare them with the bulk pressures obtained via the 
virial and compressibility routes, equations \eqref{virial equation 2D} and \eqref{compressibility pressure equation}, respectively. 
The results are shown in Figure~\ref{fig contact thm}.

The parameters of the HCY potential are selected such that the thermodynamic inconsistency between the 
virial and compressibility routes to the bulk pressure can easily be resolved. We fix $\alpha\!=\!2$ and look at two physically distinct cases, $\kappa\!=\!-1.5$, corresponding to an attractive tail, and $\kappa\!=\!10$, describing soft repulsive particles. 
We have been careful to choose the amplitude of the attractive tail in the former case such to avoid any liquid-gas phase transition 
phenomenology. This is a practical choice, since we would otherwise encounter large drying layers at the repulsive substrate which would only 
complicate the numerics of solving the inhomogeneous OZ equation and neccessitate even larger grid sizes without yielding any additional 
physical insights.
More importantly, the bulk compressibility pressure \eqref{compressibility pressure equation} is ill-defined on the liquid-side of the coexistence curve, since the required integral runs over the density-space and therefore crosses the coexistence region.

The attractive-tail case and repulsive-tail case are shown in the first and second column of Figure~\ref{fig contact thm}, respectively.
The reduced pressure curves are shown in panels A, in solid orange lines for the virial route and dashed limegreen lines for the compressibility route. The results obtained via the contact theorem are given by colored circles, red when the YBG equation was used to calculate the inhomogeneous density profile and green when the LMBW equation was used. Panels B illustrate the quantities input in the contact theorem. The density profiles generated by the LMBW equation are shown as dashed green lines, while the integrands needed in \eqref{specific wall theorem} are shown as solid green lines. (We do not show the equivalent for the YBG case, since the curves are very similar to each other; however their integrated outputs show discrepancies as can be seen in panels A.) From the results in panels A of Figure~\ref{fig contact thm} it is clear that the curves generated by the YBG equation are (as already known \cite{tschopp3}) consistent with the virial route, while the ones generated by the LMBW equation are consistent with the compressibility route. The latter result is, in retrospect, also not surprising since the LMBW equation is fully equivalent to the EL equation of standard DFT, which is known to give compressibility-consistent results.

To prove this, we start from the EL equation
\begin{equation}\label{EL equation standard DFT}
\rho(\vec{r}_1) = e^{-\beta \left( V_{\text{ext}}(\vec{r}_1) - \mu - k_B T c^{(1)}(\vec{r}_1)  \right)},
\end{equation}
where $c^{(1)}$ is the one-body direct correlation function, defined as the first functional derivative of the excess Helmholtz free energy functional with respect to the one-body density.
Taking first the logarithm and then applying a spatial gradient yields
\begin{align}\label{grad EL}
\nabla_{\vec{r}_1} \rho(\vec{r}_1) &= - \rho(\vec{r}_1) \nabla_{\vec{r}_1} \beta V_{\text{ext}}(\vec{r}_1) 
\\&\hspace*{1.8cm}
+ \rho(\vec{r}_1) \nabla_{\vec{r}_1} c^{(1)}(\vec{r}_1) . \notag
\end{align}
The gradiant of the one-body direct correlation function can be related to the two-body direct correlation function using the well-known sum-rule \cite{evans79},
\begin{equation}\label{sum-rule EL}
\nabla_{\vec{r}_1} c^{(1)}(\vec{r}_1) = \int\! d \vec{r}_2 \,  c(\vec{r}_1,\vec{r}_2) \nabla_{\vec{r}_2} 
\rho(\vec{r}_2).
\end{equation}
Substitution of equation \eqref{sum-rule EL} into the preceding equation \eqref{grad EL} we obtain
\begin{align}
\nabla_{\vec{r}_1}  \rho(\vec{r}_1) &=  -\rho(\vec{r}_1) \nabla_{\vec{r}_1} \beta V_{\text{ext}}(\vec{r}_1)
\\&\hspace*{1.8cm}
+ 
\rho(\vec{r}_1)\!\int\! d \vec{r}_2 \,  c(\vec{r}_1,\vec{r}_2) \nabla_{\vec{r}_2} 
\rho(\vec{r}_2), \notag
\end{align}
which is the LMBW equation \eqref{LMBW}.
Since this small derivation employs no relations outside the compressibility realm, we can safely conclude that the EL equation and the LMBW are formally equivalent and remain so regardless of any input approximation used (i.e.~closure relation or approximated Helmholtz free energy functional).

\subsection{Optimization by enforcing structural consistency}

If the LMBW route to the density profile is fully equivalent to the standard DFT treatment, one can ask what this (much more demanding) implementation path is useful for. Having the possibility to use integral equation closures instead of an approximate form for the Helmholtz free energy functional is desirable when treating highly packed systems of hard-core particles. Indeed, the core condition is explicitly set in closure relations (e.g. when using the Percus-Yevick closure) but not fully satisfied when using free energy functionals given by fundamental measure theory (e.g. the Rosenfeld functional for hard spheres). On the other hand, closure relations can input fully soft interaction potentials (e.g. Lennard-Jones) without the need of perturbative treatment. 
More importantly in the context of this work, having two independent routes to the one-body density profile, 
both reliant upon the same integral equation closure, triggers the possibility of using a parameterized closure 
to minimize the structural inconsistency.

\subsubsection{Packed system}

\begin{figure}[t!]
\includegraphics[width=1\linewidth]{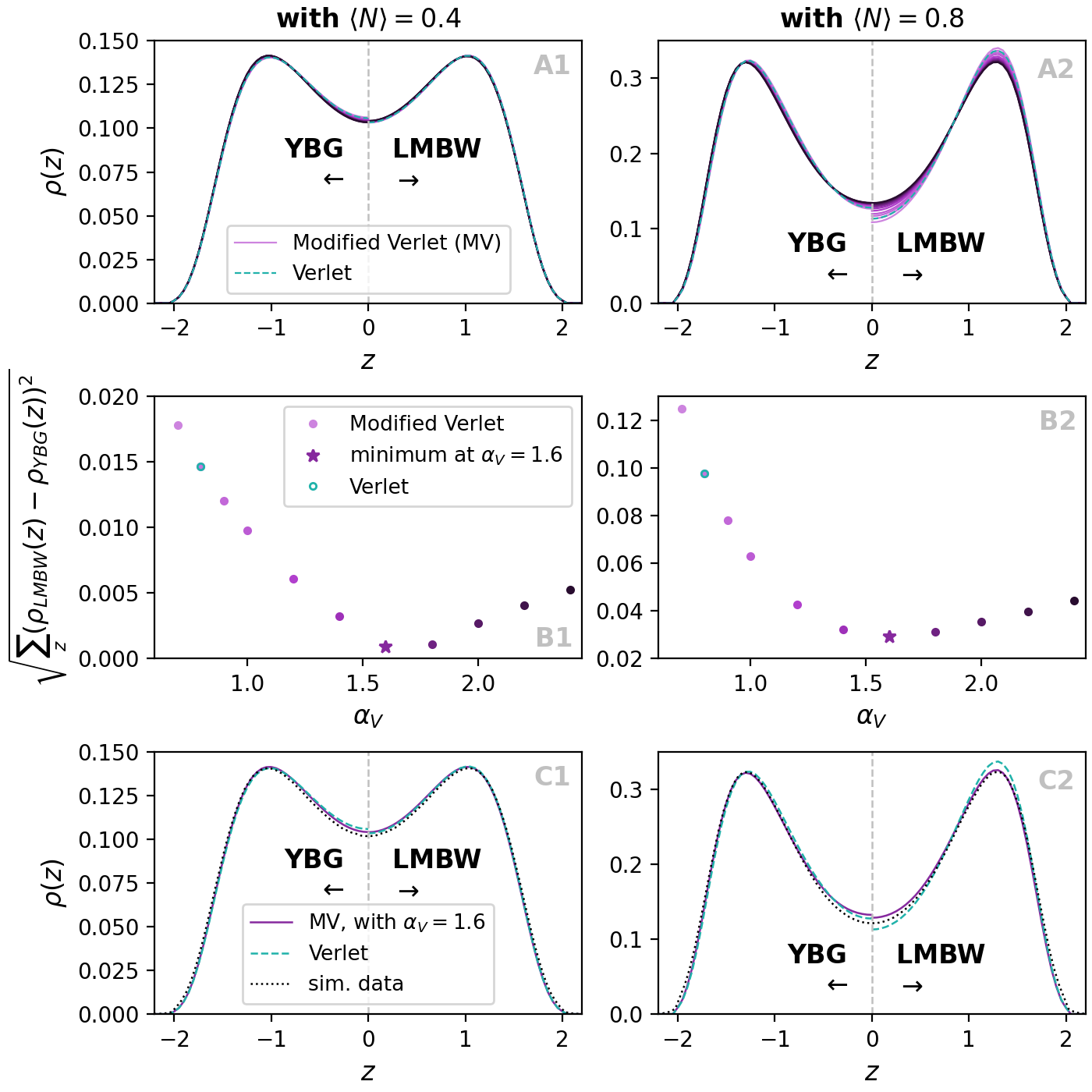}
\caption{ \textbf{Optimization of the density profiles using the Modified Verlet closure.}
In panels A we show density profiles calculated using the YBG and LMBW equations for various values of the 
optimization parameter, $\alpha_V$. Since the profiles are symmetric about $z\!=\!0$ we show both YBG and LMBW profiles on the same 
plot for ease of comparison. The left column of panels concern results for $\langle N\rangle\!=\!0.4$, while the right column 
is for $\langle N\rangle\!=\!0.8$. 
Panels B show the root-mean-square difference between the profiles obtained using the two different routes as a function of $\alpha_V$. 
The minimum of both curves is found to be at $\alpha_V\!=\!1.6$. 
Panels C show the profiles at this optimal value of $\alpha_V$ and demonstrate the improved structural consistency compared with the standard Verlet closure.
We also show simulation data as dotted black curves for comparison.
} 
\label{fig inhomo rho with MV}
\end{figure}

In Figure~\ref{fig inhomo rho with MV} we show results for the above mentioned optimization scheme applied to a system of HCY particles \eqref{HCY potential}, with parameters $\alpha\!=\!2$ and $\kappa\!=\!10$, trapped in the slit given by equation \eqref{external potential exp walls}. The columns refer to two different packing states, namley $\langle N\rangle\!=\!0.4$ for the first column and $\langle N\rangle\!=\!0.8$ for the second. Density profiles calculated using the Modified Verlet closure \eqref{MV closure} for different values of its tuning parameter, $\alpha_V$, are shown in shades of purple (with increasing value corresponding to darker shade). The density profiles being symmetric the curves for the YBG and LMBW outcomes share the same plot. As a reference, standard Verlet predictions (corresponding to the Modified Verlet closure for $\alpha_V\!=\!0.8$) are given as dashed seegreen lines. To estimate the level of structural consistency the root-mean-square difference between the YBG and LMBW profiles with respect to $\alpha_V$ is shown in panels B. The minimal structural inconsistency is reached at $\alpha_V$-value $1.6$ in both cases studied. Density profiles obtained at this optimal tuning value are shown in panels C together with simulation data (as dotted black lines) and, once more, the original standard Verlet predictions.
The optimized density profiles show indeed better structural consistency than the Verlet curves and lie closer to the simulation data.

At first sight it may be surprising that the best $\alpha_V$-value is the same for both $\langle N\rangle$-cases. However, the closure relation solely encodes the influence of the interaction potential on the pair correlations. It is thus fully independent of quantities like the external potential or the packing. 
Finding the same optimal $\alpha_V$ in both cases is therefore reassuring in its consistency with this fundamental feature. To test this further we will explore, in the following, the same system of particles both in bulk (where $V_{\text{ext}}\!=\!0$) and under a different external potential.

\subsubsection{Pressure curves}

\begin{figure}[t!]
\includegraphics[width=1\linewidth]{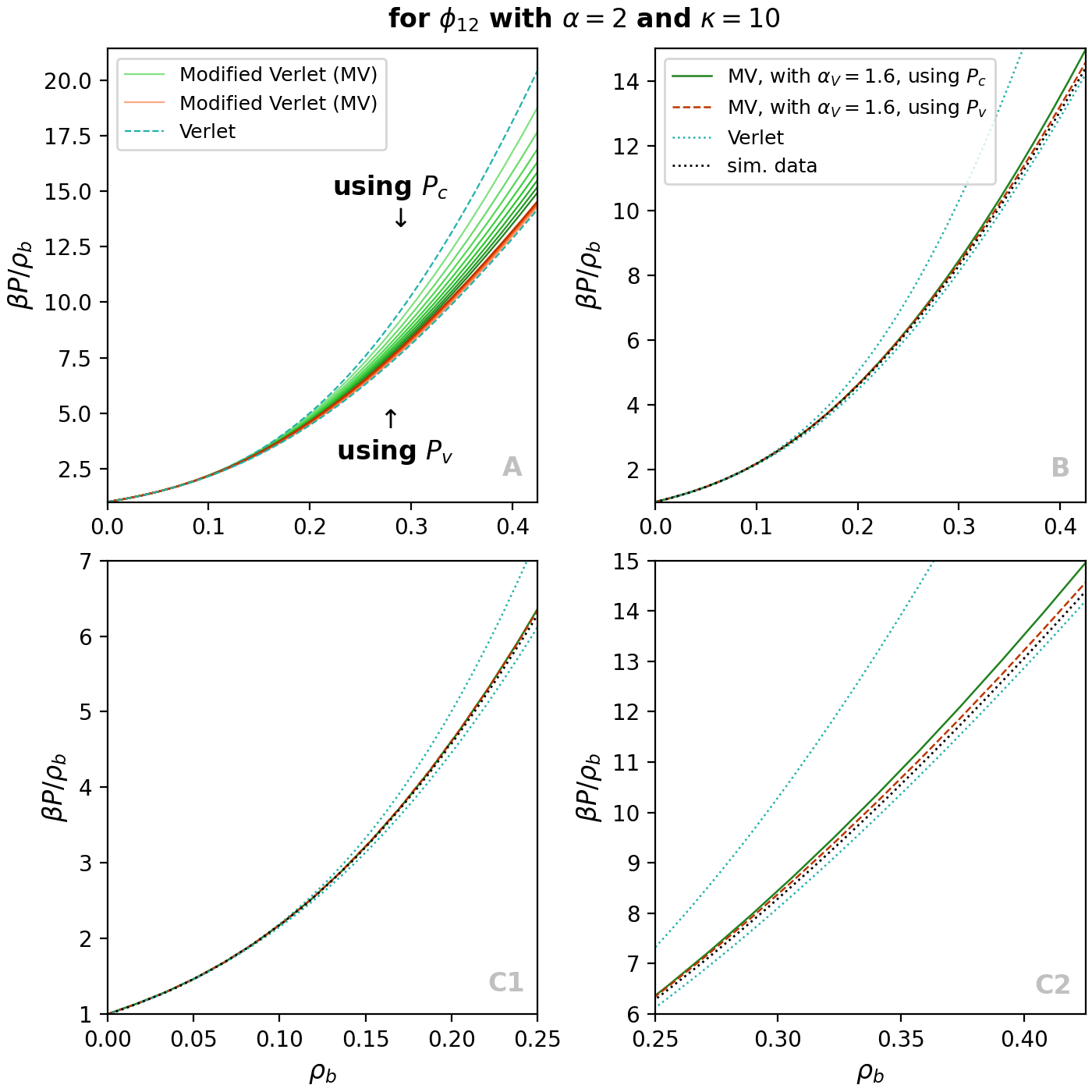}
\caption{ \textbf{Bulk pressure optimization using the Modified Verlet closure.}
In Panel A we show the bulk pressure from the virial \eqref{virial equation 2D} and compressibility \eqref{compressibility pressure equation} equations as a function of the bulk density. The dashed seagreen lines show the results obtained using the standard Verlet closure, as in Figure~\ref{fig contact thm}. Increasing the parameter $\alpha_V$ from 0.8 (the standard Verlet value) to 1.6 leads to a reduction of thermodynamic inconsistency. 
Virial pressures are shown as solid limegreen lines, while the compressibility pressures are shown in orange. 
In panel B we show only the pressures for the standard Verlet closure and the Modified Verlet closure with the optimized $\alpha_V\!=\!1.6$.
Panels C show zooms of the data from Panel B to focus on different density regimes.
We added simulation data as dotted black curves to panels B and C for comparison.
} 
\label{fig bulk pressure with MV}
\end{figure}

Let us consider once more a bulk system of fully repulsive HCY particles \eqref{HCY potential}, with parameters $\alpha\!=\!2$ and $\kappa\!=\!10$. We aim to optimize the pressure curves given by equations \eqref{virial equation 2D} and \eqref{compressibility pressure equation} for the virial and compressibility routes, respectively, by changing the Modified Verlet parameter $\alpha_V$. By repeating the same scheme as used for the inhomogeneous density profiles in the previous subsubsection, we want to ensure that the resulting optimal $\alpha_V$-value remains stable at $1.6$.

In panel A of Figure~\ref{fig bulk pressure with MV} we show the reduced pressure as a function of the bulk density. The standard Verlet results, the exact same than in panel A2 of Figure~\ref{fig contact thm}, are given as dashed seagreen lines. They correspond to the Modified Verlet closure for $\alpha_V\!=\!0.8$. The Modified Verlet closure outcomes are shown as solid lines, in shades of limegreen for the compressibility route and of orange for the virial one. As the value of the tuning parameter is increased the thermodynamic inconsistency between these two routes is reduced. The best match is reached at $\alpha_V\!=\!1.6$, which indeed agrees with the previous test-case in the inhomogeneous regime. Increasing the $\alpha_V$-value above $1.6$ generates unphysical curves that start to contradict the low-density limit.

In panel B we show the optimal Modified Verlet predictions at $\alpha_V\!=\!1.6$ together with the standard Verlet curves and simulation data (as dotted black lines). Panels C1 and C2 are zooms of panel B for both low and high density regimes, respectively.
At low density the optimised Verlet Modified curves are almost perfectly consistent with each other and with the simulation data. They represent a big improvement upon the standard Verlet predictions. For higher densities the $1.6$ Modified Verlet outcomes are much more thermodynamically consistent than the standard Verlet ones. However, some small differences between them and with the simulations become visible, which is to be expected since the Verlet Modified closure is not an exact approach and gets more severely tested with increased packing. This can be compared with column 2 of Figure~\ref{fig inhomo rho with MV} that also exhibits more deviations than the results in the first column, where the packings were $\langle N\rangle\!=\!0.8$ and $0.4$, respectively.

The consistency between the two test-cases considered so far, namely the inhomogeneous and the bulk regimes, show the stability of this optimisation scheme. We are thus confident that applying $\alpha_V\!=\!1.6$ to any other situation, as long as we do not modify the (parameters of the) interparticle interactions, would yield improved results with respect to the standard Verlet theory, with reduced structural/thermodynamic inconsistency between the routes.

\subsection{Test on a different external potential}

\begin{figure}[t!]
\includegraphics[width=1\linewidth]{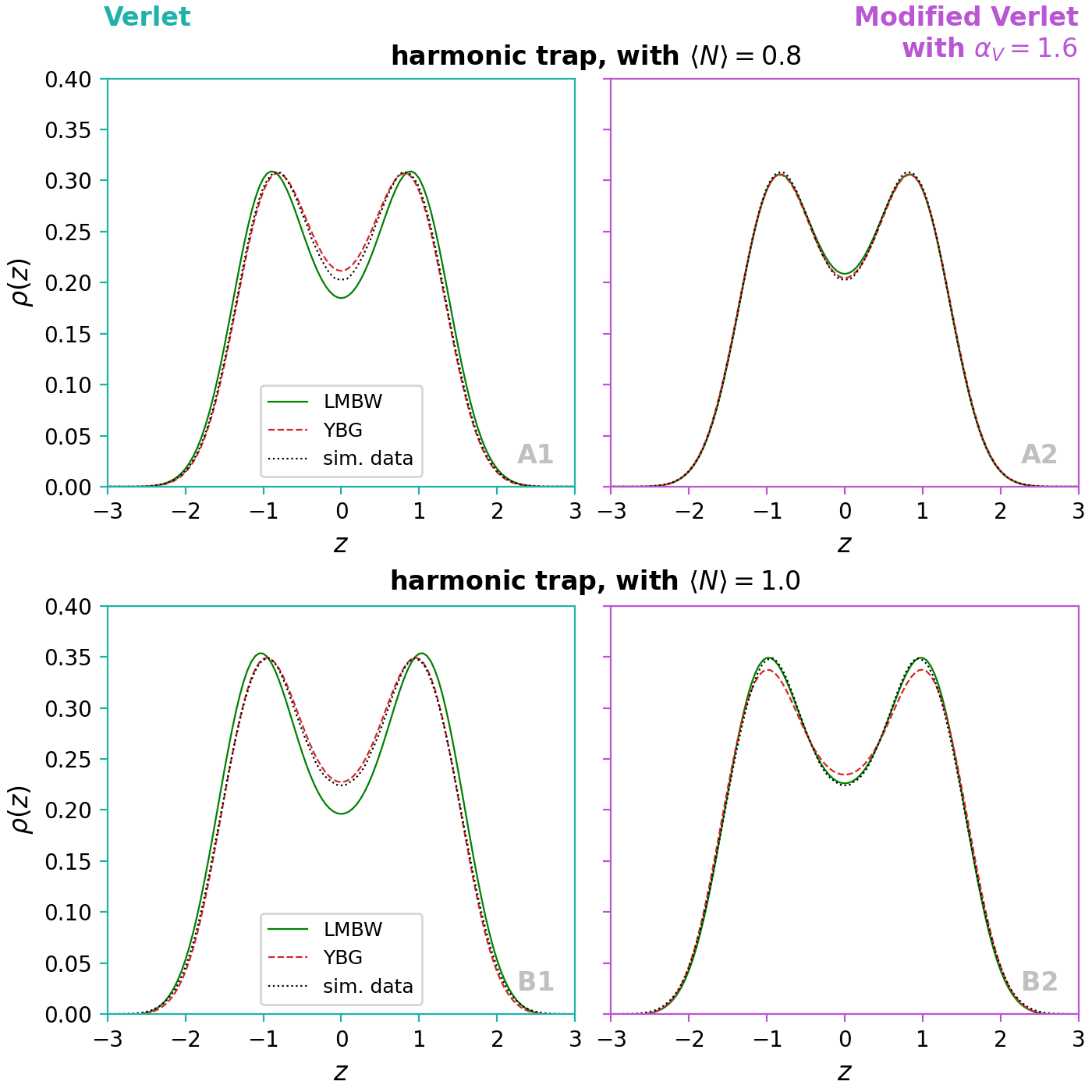}
\caption{ \textbf{Application of the optimized closure to a harmonic trap.}
We show density profiles in the external field \eqref{external potential harmonic trap}. In panels A the average number of particles per unit length is $\langle N \rangle \!=\! 0.8$, while in panels B its value is $\langle N \rangle \!=\! 1.0$. The first column (in seagreen) shows results using the standard Verlet closure. The second column (in purple) shows results using the Modified Verlet closure for fixed optimized $\alpha_V\!=\!1.6$.
The density profiles calculated with the LMBW equation are given by solid green lines and the ones calculated with the YBG equation are in dashed red. The simulation data are given by the dotted black curves.
It is clear that in both test-cases the Modified Verlet closure with $\alpha_V\!=\!1.6$ reduces the structural inconsistency in comparison to the results from the standard Verlet closure.
} 
\label{fig application other Vext}
\end{figure}

In the previous subsection we determined the optimal value of the tuning parameter in the Modified Verlet closure \eqref{MV closure}, namely $\alpha_V\!=\!1.6$, for a system of HYC particles \eqref{HCY potential}, with parameters $\alpha\!=\!2$ and $\kappa\!=\!10$. In the following we test this optimized closure on a harmonic trap given by equation \eqref{external potential harmonic trap}, which differs from the external potential used previously. The results are shown in Figure~\ref{fig application other Vext}.
The first row gives density profiles for $\langle N \rangle \!=\! 0.8$, while the second row shows results for $\langle N \rangle \!=\! 1.0$. The densities predicted by the standard Verlet closure are set in the first (seagreen) column and the predictions from the optimized Modified Verlet closure are in the second (purple) column.
For each panel we show the curves output from the LMBW equation as solid green lines, the ones from the YBG equation as dashed red lines and simulation data as dotted black lines.
In panels A, at lower packing, we observe that, indeed, the optimized closure both has much better structural consistency and agrees more faithfully with the simulation curve than the standard Verlet does.
In panels B the structural inconsistency is again strongly reduced by the optimized closure in comparison to the standard Verlet one.
It is interesting to observe that for the latter case the YBG outcomes are very close to the simulation data, while for the former it is the LMBW ones that are in strong agreement with them.
In contrast, in panels A, for the lower packing, both the standard Verlet and the optimized closure yield better results when using the YBG equation.
This simple observation illustrates the difficulty of guessing in advance which route will yield the better result.
We are thus of the opinion that enforcing the consistency first is more important than trying to guess the preferable route. When the optimized closure is employed the choice of route becomes less significant and it appears they bracket quite tightly the simulation curves.

\section{Discussion \& Conclusions}

Force-DFT, as introduced in \cite{tschopp3}, employs the YBG equation together with the inhomogeneous OZ equation to generate density profiles that are consistent with virial thermodynamics.
To enable comparison with standard DFT, known to be consistent with compressibility thermodynamics, this first implementation of force-DFT took an approximation to the excess Helmholtz free energy functional as input.
In contrast, in Reference \cite{tschopp8}, $F^{\text{exc}}$ was supplanted by an integral equation closure relation.

In the present work, we have built an alternative approach using the LMBW equation in place of the YBG. This new scheme has then been shown to be fully equivalent to standard DFT, but once more, by using a closure relation instead, remains implementable even in the absence of a known free energy functional.
We have thus been able to develop an optimized theory in which we employ the concept of structural inconsistency between 
the LMBW and YBG routes to the one-body density. 
By introducing a free parameter into the closure relation we minimize the differences between their outcomes and obtain density profiles in close agreement with simulation data.

While obtaining good agreement with simulation and making precise quantitative predictions are certaintly desirable outcomes, the integral equation approaches we have developed illuminate a fundamental aspect of liquid-state theory, namely that the closure relation, as the excess Helmholtz free energy functional, is an intrinsic property which represents the interparticle interaction solely. In our results the stability of the minimizing tuning-parameter-value aligns with that fundamental principle, by being largely indifferent to the choice of the external potential or the packing. (The former may be more obvious if one recalls that integral equation closures were originally intended for application to bulk systems.)

Now that this fundamental aspect has been explicitly tested, we can confidently use the following scheme for a given interparticle interaction: (i) find the optimal closure-parameter value that minimizes the thermodynamic inconsistency between the virial and the compressibility pressure curves (as in Figure~\ref{fig bulk pressure with MV}) and (ii) take that optimized closure relation to perform calculation in any inhomogeneous case (with force- or LMBW DFT) or even in dynamics (when using superadiabatic dynamical DFT \cite{tschopp4}).

{\red
After the initial test of the contact theorem we moved away from attractive particles and focused solely on purely repulsive interaction potentials, to avoid any issues associated with phase separation or criticality.
We anticipate that the stuctural inconsistency between the routes will be quite severely tested when considering 
an attractive system at thermodynamic statepoints close to the coexistence curve, as forshadowed by panel A1 of Figure~\ref{fig contact thm}.
Testing the performance of our optimization scheme on attractive particles, HCY or even Lennard-Jones, presents thus an interesting avenue for future investigations. 
}

In the present work we used the root-mean-square difference to assess the structural inconsistency between the YBG and LMBW density profiles. This type of measure is a natural first choice.
Potential improvements to the above presented optimization scheme could be obtained by imposing a different, and perhaps physically more meaningful, measure
{\red or by applying the same measure to a different quantity than directly to the one-body density. For example, we could
apply the root-mean-square difference to the integrand of equation \eqref{general wall theorem} which would be somewhat analogous to minimizing the difference in pressure between the two routes, on a local level. Another option would be to
use the inverse of the one-body density as a weight function and thus account for the local magnitude of the profile.}
A further possible improvement could be to use a more sophisticated parametrized closure relation, along the lines of the Rogers-Young closure \cite{rogersyoung}, which is known to perform well in bulk.


\appendix

\section{Detailed calculations to account for discontinuities}\label{appendix accounting for discontinuities}

To treat a system of hard-core particles, the YBG version of equation \eqref{EL-like equation} needs special attention when calculating $d\phi(r_{12})/d z_1$ in $A(z_1)$. The latter is namely given by
\begin{align*}
&- \int_{-\infty}^{+\infty} \!\! dx_2\int_{-\infty}^{+\infty} \!\! dz_2 \, \rho^{(2)}(z_1,x_2,z_2 ) \frac{d\phi(r_{12})}{d z_1} \\
&= - \int_{-\infty}^{+\infty} \!\! dx_2\int_{-\infty}^{+\infty} \!\! dz_2 \, \rho^{(2)}(z_1,x_2,z_2 ) \frac{(z_1-z_2)}{r_{12}} \frac{d\phi(r_{12})}{d r_{12}}\\
&= \int_{-\infty}^{+\infty} \!\! dx_2\int_{-\infty}^{+\infty} \!\! dz_2 \, \rho^{(2)}(z_1,x_2,z_2 ) \frac{(z_1-z_2)}{r_{12}} \delta(r_{12}-d) \\
&= \int_{-\infty}^{+\infty} \!\! dr_{12}\int_{-\infty}^{+\infty} \!\! dz_2 \,\rho^{(2)}(z_1,x_2(r_{12}),z_2 ) \frac{(z_1-z_2)}{x_2(r_{12})} 
\\&\hspace*{6.35cm} \times 
\delta(r_{12}-d) \\
&= \int_{z_1-d}^{z_1+d} \!\! dz_2 \, \rho^{(2)}(z_1,x_2^\star,z_2 ) \frac{(z_1-z_2)}{\sqrt{d^2-(z_1-z_2)^2}}\\
&= \int_{z_1-d}^{z_1+d} \!\! dz_2 \, \rho^{(2)}(z_1,x_2^\star,z_2 ) \left(\frac{d}{dz_2}\sqrt{d^2-(z_1-z_2)^2}\right)\\
&= - \int_{z_1-d}^{z_1+d} \!\! dz_2 \, \sqrt{d^2-(z_1-z_2)^2} \left(\frac{d}{dz_2}\rho^{(2)}(z_1,x_2^\star,z_2 )\right) ,\\
\end{align*}
where $x_2^\star\!\equiv\!\sqrt{d^2-(z_1-z_2)^2}$ and where we have used that $x_2(r_{12})\!=\!\sqrt{r_{12}^2-(z_1-z_2)^2}$, since the distance between the center of the particles is $r_{12}\!=\!\sqrt{x_2^2 + (z_1-z_2)^2}$.

When using hard walls as an external potential, the LMBW version of equation \eqref{EL-like equation} needs special care when calculating $d \rho(z_2)/dz_2$ in $A(z_1)$.
For a slit, the integral over $z_2$ is given by
\begin{align*}
&\int_{-\infty}^{+\infty} \!\!\! dz_2 \, c(z_1,x_2,z_2 ) \frac{d \rho(z_2)}{d z_2}\\
&= \int_{z_{\text{wall}}^{\text{left}}}^{z_{\text{wall}}^{\text{right}}} \!\!\! dz_2 \, c(z_1,x_2,z_2 ) \frac{d \rho(z_2)}{d z_2}
\\&\quad+ \int_{-\infty}^{+\infty} \!\!\! dz_2 \, c(z_1,x_2,z_2 ) \rho(z_2)  \delta{(z_2-z_{\text{wall}}^{\text{left}})}
\\&\quad- \int_{-\infty}^{+\infty} \!\!\! dz_2 \, c(z_1,x_2,z_2 ) \rho(z_2)  \delta{(z_2-z_{\text{wall}}^{\text{right}})}\\
&= \int_{z_{\text{wall}}^{\text{left}}}^{z_{\text{wall}}^{\text{right}}} \!\!\! dz_2 \, c(z_1,x_2,z_2 ) \frac{d \rho(z_2)}{d z_2}
\\&\quad+ {\big(c(z_1,x_2,z_2 ) \rho(z_2) \big)}\big\rvert_{z_2=z_{\text{wall}}^{\text{left}}}
\\&\quad- {\big(c(z_1,x_2,z_2 ) \rho(z_2) \big)}\big\rvert_{z_2=z_{\text{wall}}^{\text{right}}} \; . \\
\end{align*}

\section{Details on the simulation data curves}\label{appendix sim data details}

Here below, we give the details on the simulation curves shown in Figures~\ref{fig inhomo rho with MV}, \ref{fig bulk pressure with MV} and \ref{fig application other Vext}.

\subsection{Packed system}

Following the protocol of subsection~\ref{sim data main}, we perform two simulation sets for each value of $\langle N\rangle$: 
(i) a baseline set using solely the finite part of the external potential \eqref{external potential exp walls}, and 
(ii) an augmented set that adds hyperbolic tangent functions to it, such to impose a hard, but continuous, repulsion when the particle centers exceed $|z|>2$, i.e.
\begin{align*}
V_{\mathrm{wall}}(z) &= 5\Bigl(\,1+\tanh\!\bigl(600\,(z-2)\bigr)\Bigr)\\
&\quad +5\Bigl(\,1+\tanh\!\bigl(600\,(-z-2)\bigr)\Bigr).
\end{align*}
We calculate the density profile, $\rho(z)$. For each trajectory, we enforce the known mirror symmetry by setting
\begin{equation*}
\rho_{\mathrm{sym}}(z) = \tfrac{1}{2}\,\bigl(\rho(z)+\rho(-z)\bigr).
\end{equation*}
We then average the resulting symmetric profiles over the two simulation sets (baseline and augmented) to obtain the final curves reported in Figure~\ref{fig inhomo rho with MV}.

\subsection{Pressure curve}

We perform BD simulations of HCY particles in a periodic square box of size $L_x \times L_z = 200d \times 200d$ at varying bulk number densities, $\rho_b = N/A$ (with $A=L_xL_z$). For each state point, the instantaneous mechanical pressure is calculated as the sum of the ideal (osmotic) and interaction (virial) contributions via the pressure tensor
\begin{equation*}
\Pi_{\alpha\beta}(t)
= \rho\,k_{\mathrm B}T\,\delta_{\alpha\beta}
+\frac{1}{A}\sum_{i<j} r_{ij,\alpha}\,f_{ij,\beta},
\end{equation*}
in reduced units ($k_{\mathrm B}T\!=\!1$) and where $\alpha,\beta\in\{x,z\}$ label Cartesian components in two dimensions, $\mathbf r_{ij}\!=\!\mathbf r_i-\mathbf r_j$ is the minimum-image separation vector with components $r_{ij,\alpha}$ and $\mathbf f_{ij}$ is the force on particle $i$ due to particle $j$, with components $f_{ij,\beta}$.
In two dimensions, the scalar pressure is $P(t)=\tfrac{1}{2}\,\mathrm{Tr}\,\Pi(t)$. At each density, the reported pressure is obtained by time-averaging $P(t)$ over the production window of the simulations.
The final curve is shown in Figure~\ref{fig bulk pressure with MV}.

\subsection{Test on a different external potential}
We calculate density profiles from BD simulations of HCY particles using the finite part of the alternative external potential defined in equation \eqref{external potential harmonic trap}, while keeping all other model parameters and the simulation protocol identical to those in subsection~\ref{sim data main}.
(Note that, since the present harmonic external potential is more constraining than the previously explored exponential walls potential, there was no need for generating a second augmented set and we relied solely on the baseline set.)
The outcomes are shown in Figure~\ref{fig application other Vext}.

\bibliography{paper_LMBW_YBG}

\end{document}